\documentclass[conference,10pt]{IEEEtran}
\usepackage{amsmath}
\usepackage{amssymb}
\usepackage{amsfonts}
\usepackage{graphicx}
\usepackage{enumerate}
\usepackage{bbm}

\usepackage{mathabx} 
\usepackage{mathtools} 
\usepackage{cite}
\usepackage{subfigure}
\usepackage{url}
\usepackage{array}
\usepackage{verbatim}
\usepackage{comment}
\usepackage{stfloats}

\usepackage{color}
\definecolor{red}{rgb}{1,0,0}
\definecolor{blue}{rgb}{0,0,1}

\long\def\comment#1{}

\newfont{\bbb}{msbm10 scaled 700}

\newfont{\bb}{msbm10 scaled 1100}

\newcommand{\Cm}{{\bf C}}

\newcommand{\Gm}{{\bf G}}

\newcommand{\Xm}{{\bf X}}

\newcommand{\Dc}{{\cal D}}

\newcommand{\Rc}{{\cal R}}

\newcommand{\Uc}{{\cal U}}

\newcommand{\Vc}{{\cal V}}

\newcommand{\Thetam}{\hbox{\boldmath$\Theta$}}

\newcommand{\eqdef}{\stackrel{\Delta}{=}}

\begin{document}
\title{ 
Efficient C-RAN Random Access for IoT Devices: Learning Links via Recommendation Systems}

\author{\IEEEauthorblockN{Ozgun Y. Bursalioglu$^1$, Zheda Li$^2$, Chenwei Wang$^1$, Haralabos Papadopoulos$^1$}
\IEEEauthorblockA{$\ ^1$Docomo Innovations Inc., Palo Alto, CA\\
$\ ^2$Dept. of EE, University of Southern California, Los Angeles \\
\{obursalioglu, cwang, hpapadopoulos\}@docomoinnovations.com, zhedali@usc.edu}}

\maketitle
\begin{abstract}
%

We focus on C-RAN random access protocols for IoT devices that yield low-latency high-rate active-device detection in dense networks of large-array remote radio heads. In this context,  we study the problem of learning the strengths of links between detected devices and network sites. In particular, we develop recommendation-system inspired algorithms, which exploit  random-access observations collected across the network to classify links between active devices and network sites across the network. Our simulations and analysis reveal the potential merit of data-driven schemes for such on-the-fly link classification and  subsequent resource allocation across a wide-area network. 
%
\end{abstract}



\vspace{-0.1cm}
\section{Introduction}
\label{section:introduction}
\vspace{-0.15cm}

A key challenge for 5G and beyond-5G deployments will be, without doubt, a seamless integration of Internet of Things (IoT) services, as these are expected to span a very broad range of application scenarios and technical requirements. The envisioned high IoT device diversity and density  make the problem  even more challenging.  It is also expected that large antenna arrays, massive MIMO, and the centralized radio access network (C-RAN) architecture will be an integral part of  future deployments and can be exploited to enable efficient network access to IoT devices.

 


While future networks may need to accommodate  humongous numbers of IoT devices, the  communication requirements  of most IoT devices are expected to be intermittent and sporadic.  As a result, random access (RA) protocols are ideally suited for serving IoT devices.  Most existing RA works consider cellular RA, i.e., access to a single-site \cite{codedAlohavsCS2017, 7-Larsson}.  RA protocols for C-RAN architectures,  however, are also gaining attention \cite{xu-cran-icc2015,qihe-cran-2017}.

Recent advances in cellular RA include coded and slotted versions of ALOHA, and compressed sensing (CS) techniques  aiming to exploit the inherent sparsity in user activity (see e.g., \cite{codedAlohavsCS2017}).   Some of the benefits offered by massive MIMO have already been exploited
in cellular RA, and include
 spatial collision resolution (see e.g., \cite{7-Larsson}). 

Several C-RAN RA methods have emerged recently.
\cite{xu-cran-icc2015} leverages  CS for sparse user detection, and clustering of nearby base stations (BS)  to reduce computational complexity. \cite{qihe-cran-2017} also exploits  CS techniques for user detection, but, unlike \cite{xu-cran-icc2015}, it performs user detection and channel estimation jointly by modeling a two-dimensional sparsity, i.e., sparsity in both device activity and device-site connectivity. While the two-dimensional sparsity model is well-suited to C-RAN architectures with IoT devices, the computational complexity in \cite{qihe-cran-2017} scales up with the number of antennas at both BSs and users. 

Contention-based  cellular RA is preferred in LTE as a means for accommodating IoT devices with  sparse and intermittent activity \cite{good_lte-explantion}. Uplink (UL) pilot collisions, detected by a BS, are resolved using additional message exchanges between BS and users \cite{good_lte-explantion}.  The inherent delays in resolving collisions reduce the appeal of these schemes, especially in the presence of networks with small cells and high IoT densities.

In this work, we consider C-RAN RA protocols that enable high detection rates of  active devices  in the network and study the problem of learning the large-scale channel gains of the links between detected devices and C-RAN sites.  
Such link information can be leveraged for scheduling, load-balancing,  and interference suppression to greatly improve performance in dense massive MIMO deployments \cite{bethanabhotla2016optimal}, \cite{qiaoyang}.

We focus on a user-centric architecture of the type introduced in \cite{bursalioglu2016novel} that
 leverages dense deployments of large-array remote radio heads (RRHs)  and virtual sectorization.   This architecture is ideal for IoT RA, as it allows on-the-fly low-latency detection of active devices across the C-RAN coverage area. It also greatly outperforms conventional cellular RA  in terms of density of simultaneously detected devices  \cite{bursalioglu2016novel,icc2016-ozgun}.
 
In the context of these RA protocols, we present techniques for learning the strengths of links between detected devices and RRH sectors in the network. By jointly processing 
the individual RRH sector device-detection reports via a framework borrowed from recommendation systems \cite{netflix-winner-2009}, the proposed methods are able to  effectively classify the strengths of links between detected devices and RRH sectors not directly available in the  detection reports.  Our proposed schemes only exploit prior knowledge of the distances between RRH sites, but no knowledge of active-device locations.

%

 The paper is organized as follows. Sec.~\ref{sec:RA} describes the C-RAN RA schemes  of interest. The problem of  learning the strengths of as many  links as possible is considered in detail in Sec. \ref{link-detection}. Sec. \ref{Numerical-Results} presents a brief simulation-based evaluation of the link classification schemes proposed in Sec. \ref{link-detection}. Finally 	Sec.~\ref{sec:conclusion} provides concluding remarks and directions for future investigations.
 \vspace{-0.3cm}
\section{C-RAN Random Access} \label{sec:RA}
In this section we present the RA schemes of interest in this work. 
We consider  a C-RAN of $B$ large antenna-array BSs 
spanning a wide geographical area and serving a very large number of IoT devices, each with a  single antenna.
Similar to \cite{Li2016, bursalioglu2016novel}, spatial processing is applied at each BS to create $S$ virtual sectors, for a total of $V = SB$ sectors in the network.  The set of all BS sectors  is indexed via  the index $v=Sb+s \in \Vc = \{0,\,1,\,\cdots,\,V-1\}$, where $b \in \{0,\,1,\,\cdots,\,B-1\}$ is the BS index, and $s\in \{0,\,1,\,\cdots,\,S-1\}$.

\begin{figure}
\centering
\includegraphics[width=8.2cm, height = 3.3cm]{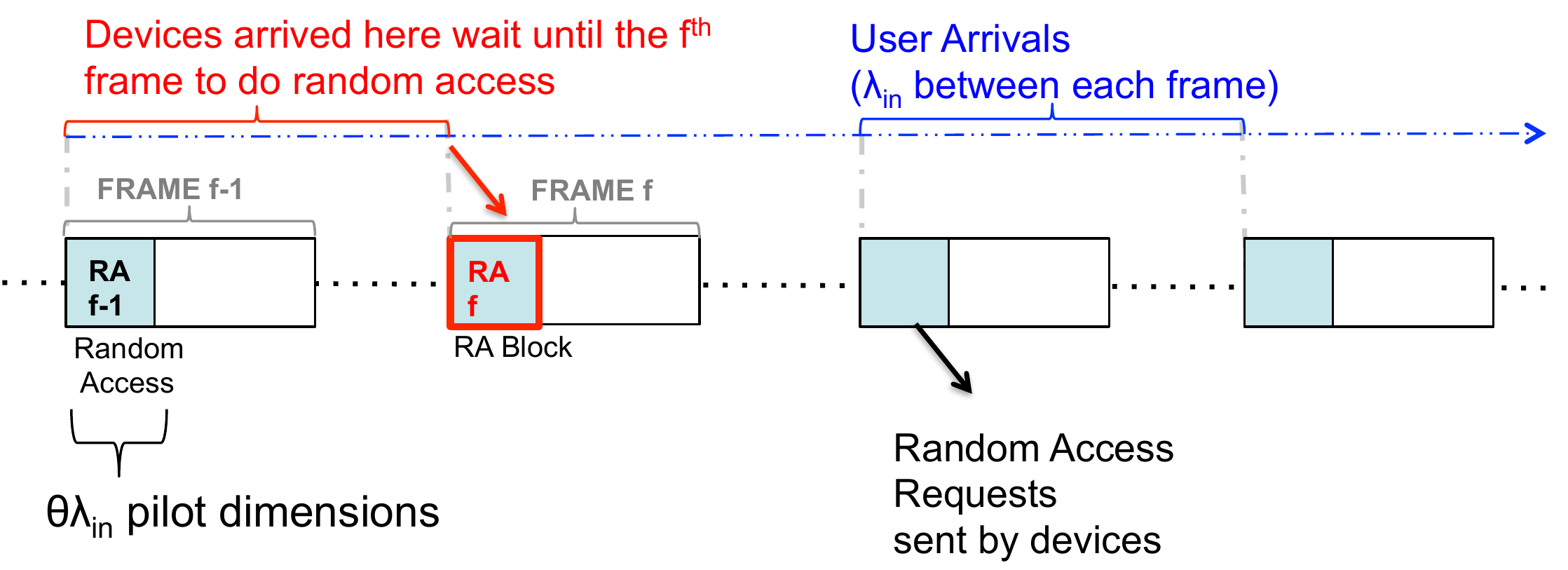}
\vspace*{-0.2cm}
\caption{Slotted RA: RA frames and blocks}
\label{fig:frame}

\end{figure}

We consider a slotted RA scheme, according to which blocks of time-frequency resources, referred to as RA blocks,  are reserved for RA \cite{good_lte-explantion}.
For convenience, we consider a timeline where RA blocks are periodically reserved. 
A single frame is formed by one RA block and the following block of resources dedicated to serving the active devices detected during the RA block in the frame. The RA block within the $f$-th frame is referred to as the $f$-th RA block. See Fig. \ref{fig:frame}.

We model the  device access-request arrival process as a Poisson process with a rate of $\lambda_{\rm in}$ device requests between consecutive RA blocks (over the geographical area spanned by the network). 
Any device with an ``arriving'' access request waits for the next RA block to access the network and remains ``active'' until it is detected. It is served in the serving block following the RA-block it is detected. 
We also let $\Rc_{f}$ denote the active users in the $f$-th frame.

During the $f$-th RA block,  active devices in $\Rc_{f}$ broadcast uniquely distinguishable uplink pilots and a subset of them, $\Dc_f$, is detected by the network.
In the C-RAN RA scheme we consider, an active user is considered to be detected by the network as long as at least one sector detects its unique ID during RA. 

Much like any RA protocol, not all active devices in an RA block are detected.  The subset of active devices, $\Rc_{f}-\Dc_f$  that are not detected are ``queued", i.e., continue to be active in later RA-blocks. Thus, the active devices in frame $f$, either arrived in frame $f-1$, or in an earlier frame but have not yet been detected. 

An important property of a good RA protocol is keeping stable queue sizes. By letting
$\lambda_{\rm out}\eqdef \lim_{N \rightarrow\infty}\frac{1}{N}\sum_{f=1}^N|{\Dc}_f|$ denote the device detection rate, it is important that the RA scheme is stable with respect to $\lambda_{\rm in}$, that is,   $\lambda_{\rm out}=\lambda_{\rm in} $.  This implies that the active user rate per frame, $\lambda_{\rm ra}\eqdef \lim_{N \rightarrow\infty}\frac{1}{N}\sum_{f=1}^N|{\Rc}_f|$ does not diverge.

For a stable RA protocol, the  ratio  $\lambda_{\rm in}/\lambda_{\rm ra}$ is inversely proportional to the expected delay (in RA frames) that a device waits before it is detected. 
A desired average delay of $1/\rho$ can be achieved if, $\lambda_{\rm in}$ (and $\lambda_{\rm out}=\lambda_{\rm in}$) is greater than $\rho\lambda_{\rm ra}$. Stable RA protocols that yield short  expected detection delays while keeping RA-block overheads low are highly desirable.\footnote{In this work,  the RA-block overheads are proportional to the number of random-access slots, or ``pilot dimensions'', reserved in an RA-block.}

At this stage it is worth contrasting the slotted RA scheme in consideration against LTE-PRACH. LTE employs a form of slotted RA  that can be modeled at an abstract  level via the schematic in Fig.~\ref{fig:frame}. However, each device in LTE is associated with a single BS, and therefore, PRACH is designed for RA between devices and a single BS. To enable such cellular RA, resource reuse is employed within the RA block in LTE, so that the RA-block pilot dimensions used within a cell are not available in nearby cells. That loss in overhead efficiency in LTE is well known. In practice, it is alleviated by using partially overlapped pilot resources in nearby cells, which inherently results in intercell RA-block interference.  An additional difference arises from the fact that, in LTE, collisions are resolved via a sequence of message exchanges between a
BS and its colliding devices. In contrast, the C-RAN RA schemes we consider rely on instantaneous user identification and collision detection, thereby resulting in lower detection delays and lower protocol  complexity.

The proposed C-RAN slotted RA enables very flexible operation in the serving block as it enables instantaneous association of the detected devices to one or multiple BS sectors in the network.  Efficient network operation, whether cellular, CoMP (coordinated multipoint), 
or cell-free, 
can be enabled if channel conditions between detected devices and nearby sectors are available to the network. As an example,  in massive MIMO deployments, network performance can be greatly improved via optimized scheduling, load-balancing, and interference suppression,  if the large-scale channel gains (link gains) between devices and BS sectors are available \cite{qiaoyang, bethanabhotla2016optimal}. 
 In Sec. \ref{specific-RA}, we describe the specific C-RAN RA protocol  we consider for detecting devices.
 It is based on an adaptation of coded pilot designs of  \cite{icc2016-ozgun} for RA.\footnote{The  collision detection capability of these codes for RA is also advocated in \cite{7-Larsson}.}
 Assuming the RA protocol of Sec. \ref{specific-RA}, in Sec. \ref{link-detection} we consider the problem of effectively classifying the strengths of links  between  detected devices and BS sectors in an effort to enable efficient C-RAN operation during the serving blocks in each frame.

\subsection{RA protocol based on coded pilot designs of \cite{icc2016-ozgun} }
\label{specific-RA}
In this section, we describe the protocol according to which active devices engage  in RA along with the mechanisms used at the network side to detect active devices. We focus on a fixed but arbitrary RA frame, $f$, and, for convenience, omit the dependence of all parameters and variables on the RA frame index, $f$. We assume that the link gain between any sector $v$ and any device $u$, $g_{vu}$, stays constant for the duration of the frame.  We classify the link between sector $v$ and device $u$, as strong when the link strength exceeds a predetermined  (operator chosen)  $\Gamma$, and as weak otherwise.

 In the protocol we consider, each RA block comprises $T$ RA pilot slots during which each active device   in $\Rc$ can access the medium. 
In any pilot slot, each active device accesses the medium with probability $p$. In particular, the pilot activity of device $j \in \Rc$ in pilot slot $t$ is captured by a Bernoulli($p$) random variable, $\chi_{j}(t)$. The $\chi_{j}(t)$'s are statistically independent in $j$ and $t$.
The RA overhead parameter $T$ and the access probability $p$ are chosen  by the network operator, and are assumed fixed and known to all the devices in the network.

The RA  protocol we consider leverages the UL pilot designs considered in \cite{icc2016-ozgun} together with large-antenna arrays at each BS.  
Each device is a priori statically assigned a unique pilot codeword that the device uses in any slot that it sends a pilot.  By appropriately processing its pilot-slot observations through its large array, each BS is able to obtain the IDs of any active device that is the only one with a strong link to one of its sectors among all the active devices broadcasting a pilot in a given pilot slot  \cite{icc2016-ozgun}. In particular, the combination of the codes  in  \cite{icc2016-ozgun} and the BS processing enable a BS sector to gather the following information during a fixed but arbitrary pilot slot:
\begin{itemize}
\item[(i)] when multiple active devices with strong links to the BS sector  transmit pilots in the slot, the BS sector detects a {\em collision};
\item[(ii)] when no active devices with strong links to the BS sector  transmit pilots in the slot, the BS sector detects {\em silence};
\item[(iii)] when a single active device with a strong link to the BS sector  transmits a pilot in the slot, the sector detects the user ID and obtains an accurate estimate of the link gain.
\end{itemize}
%
%
Each BS sector provides the information it has gathered through its $T$ pilot slots to the central unit.
By merging this information the central unit is able to obtain  $\Dc$, the  set of detected active devices across the whole network. It is also able to obtain additional valuable information, including: (i) $\Vc_{t}^{\emptyset}$:  the set of sectors that detected silence on pilot dimension $t$; (ii) $\Vc^{j}$: the set of sectors that detected the $j$-th active device (assuming all active devices in $\Dc$ are re-indexed from 1 to $|\Dc|$); (iii) $\Vc_{t}^{-, j}$: the set of sectors that on pilot dimension $t$ detected a device {\em different} from $j$; (iv) link gains between BS sectors and the active devices they detected.

\section{Classifying the active device links}
\label{link-detection}


In this section we focus on classifying the strengths of as many links between active devices  and BS sectors as possible based on the information provided in the  RA reports of all the BS sectors.
In particular, focusing on a fixed but arbitrary RA frame, the goal  is  to classify at the central unit the links between the  BS sectors and the detected devices into two classes: strong and weak.  Each  BS sector's RA report contains $T$ entries, one per pilot dimension.  In accordance to the RA protocol in Sec.~\ref{sec:RA}, for each $t \in \{1,\,2, \,\cdots,\,T\}$, each sector reports one of the following: (i) a pilot collision; (ii) pilot silence; (iii) device detection, along with the identity of the detected device and the associated link gain. 

%

In Sec. \ref{problem-formulation} we formulate the strong link detection problem of interest. In Sec. \ref{baseline} we present a baseline scheme that classifies link strengths by processing the RA report information per detected device (i.e., individually). Specifically,  the central unit processes the sector  RA reports separately per device, and, in the process, infers all the information it can regarding links between the given device and BS-sectors. In Sec. \ref{matrix-completion}, we develop a class of methods motivated by what are referred to as recommendation systems. These online methods jointly classify the links of all detected devices across the network,  exploiting in the process the inherent spatial correlation in the link strengths of nearby detected devices without making use of device location information.
\subsection{Problem Formulation}
\label{problem-formulation}

We focus on the problem of classifying the strengths of all links in the network (i.e., between all detected devices and all BS sectors) as strong and weak, during a fixed but arbitrary RA frame $f$ and omit the dependence of all parameters and variables on the RA frame index, $f$. 
The link gains between the $V$  BS sectors  and the set of $\Dc$ detected devices in the frame are compactly represented via   a $V \times |\Dc|$ matrix $\Gm$. By letting $o_j\in \Dc$ denote the ID of the $j$-th detected device in the frame, the $(i,j)$-th entry of $\Gm$ contains the link gain between the $i$-th BS sector and the $j$-th detected device, i.e.,  $\Gm_{i,j}= g_{i,o_j}$. 

Links are classified as weak or strong based on the comparison against a predefined classification threshold $\Gamma$.  To this end, we refer to the $V \times |\Dc|$ matrix $\Cm$,  with $(i,j)$-th entry 
\begin{equation}
\label{Cmdef}
\Cm_{i,j} =\begin{cases}
1 & \text{if $\Gm_{i,j}\geq \Gamma$}, \\
0 & \text{otherwise}
\end{cases} \ 
\end{equation}
as the hypothesis matrix.  The task of interest amounts to forming a classifier $\widehat{\Cm}$ at the central control unit using the BS sector RA reports, so that $\widehat{\Cm}$ is as close to  $\Cm$ as possible.
%

The classification performance is measured in terms of  probability of 
detection and probability of false alarm, {\em viz.}, 
\begin{subequations}
\label{PDPFdef}
\begin{eqnarray}
 \label{PDdef}
P_{\rm  D} & =& {\rm Pr}\left(\widehat{\Cm}_{i,j} = 1|{\Cm}_{i,j} = 1\right),\\
 \label{PFdef}
 P_{\rm  F} & =&  {\rm Pr}\left(\widehat{\Cm}_{i,j} = 1|{\Cm}_{i,j} = 0\right)\ .
\end{eqnarray}
\end{subequations}
Our performance evaluations rely on estimates of the quantities in (\ref{PDPFdef}) obtained over $N$ RA frame realizations with large $N$:
\begin{subequations}
\begin{eqnarray}
\widehat{P}_{\rm  D} &=&  
\frac{1}{N}\sum_{f=1}^N \frac{\sum_{i,j} 1_{\{\widehat{\Cm}(f)_{i,j} = 1,{\Cm}(f)_{i,j} = 1 \}}}{\sum_{i,j} 1_{{\{\Cm}(f)_{i,j} = 1 \}}}, \\
\widehat{P}_{\rm  F} &=&
\frac{1}{N}\sum_{f=1}^N \frac{\sum_{i,j} 1_{\{\widehat{\Cm}(f)_{i,j} = 1,{\Cm}(f)_{i,j} = 0 \}}}{\sum_{i,j} 1_{{\{\Cm}(f)_{i,j} = 0 \}}}\ ,
\end{eqnarray}
\end{subequations}
and where we made the dependence of $\widehat{\Cm}$ and ${\Cm}$ on the RA frame index, $f$, explicit.

%

\subsection{Baseline Scheme}
\label{baseline}

In this section, we present a baseline scheme, which obtains an estimate $\widehat{\Cm}$ by processing the RA reports {\em separately} per detected device.  The baseline classifier exploits the fact that some entries of $\Cm$ are directly available or can be indirectly inferred from the BS-sector RA reports. 
Let $\Omega$ denote the subset of $(i,\,j)$ entries of the matrix $\Cm$ (i.e., subset of links) where the value $\Cm_{i,j}$ is known to the central controller. Also, let $\widebar\Omega$ denote the complement of $\Omega$, i.e., the set comprising of all $(i,\,j)$ entries for which $\Cm_{i,j}$ is not known to the controller.

Naturally, the baseline scheme sets $\widehat{\Cm}_{i,j}=\Cm_{i,j}$ for all $(i,\,j)\in \Omega$. 
Furthermore, the baseline scheme applies a randomized decision to determine the value of $\widehat{\Cm}_{i,j}$ for each  $(i,\,j)\in \widebar\Omega$.  In particular, for  each $(i,j)\in \widebar\Omega$ the baseline scheme flips a biased coin whose probability of heads equals $\alpha$. The scheme sets $\widehat\Cm_{i,j}=1$ if the coin-flip outcome is heads, and sets $\widehat\Cm_{i,j}=0$, otherwise. In summary,
 \begin{equation}
\widehat\Cm_{i,j}=\begin{cases}
\Cm_{i,j}& \text{if}\;(i,j)\in\Omega, \\
1 &  \text{if $(i,j)\in\widebar\Omega$, and $c_{i,j}$ = ``heads''}  \\
0 & \text{if $(i,j)\in\widebar\Omega$, and $c_{i,j}$  = ``tails''}
\end{cases}, 
\end{equation}
and where $c_{i,j}$  denotes the $\alpha$-biased coin-flip outcome associated with entry $(i,j)$.

Due to the randomization in  the baseline scheme, its Receiver Operating Characteristic (ROC) curve is piecewise linear.   With $\alpha = 1$, all links in $\widebar\Omega$ are classified as strong, yielding the point in the ROC with the maximum false alarm and detection rates  $(P_{\rm  F,max},P_{\rm  D,max}=1)$. For the scheme with $\alpha = 0$, all links in $\widebar\Omega$ are classified as weak,  yielding the point in the ROC with the minimum false alarm and detection rates $(P_{\rm  F,min}=0,P_{\rm  D,min})$. Varying the value of $\alpha$ in the range $[0,1]$ yields all the $(P_{\rm  F},P_{\rm  D})$ points on the line segment connecting $(P_{\rm  F,min},P_{\rm  D,min})$ and $(P_{\rm  F,max},P_{\rm  D,max})$.

We next focus on determining the set $\Omega$, or equivalently the set of links that can be obtained from the BS-sector RA reports.  It is convenient to express $\Omega$ as  $\Omega = \Omega^1 \cup  \Omega^0 $,
with $\Omega^1$ ($\Omega^0$) denoting the sets of locations of the 1's (0's) in $\Cm$  that can be obtained from these reports. 

{\bf Obtaining the known strong links from the RA reports:}
Some entries of $\Cm$ are directly available from the RA reports. 
Focusing, in particular, on the $j$-th detected device, for an arbitrary but fixed index $j$, we note that $\Cm_{i,j}$ is available to the controller for all sectors $i$ that detected device $j$. 
By letting 
\[
\Omega^1_j=\{ (i,j):  \; \text {$i\in \Vc^j$}\},
\]
with $\Vc^{j}$ denoting the BS sectors that  have detected the $j$-th device, 
all the 1 entries in $\Cm$  with coordinates in $\Omega^1= \bigcup_{j=1}^{|\Dc|} \Omega^1_j$ are  directly available at (known to) the central controller.

{\bf Inferring zeros:}
By preprocessing  the RA report per device, $j$, the baseline scheme is also able to infer {\em additional} entries of $\Cm$, and, in particular, zero entries (i.e, weak links). 

First, consider any pilot dimension during which the $j$-th detected device has been detected by a sector in the network: 
Any sectors that detected silence or a different device (from $j$) in the same pilot dimension {\em must} have a weak link to the $j$-th device.  This implies that the following set of entries (all along the $j$-th column) of $\Cm$ are known to be zero:
\begin{equation}
\widetilde \Omega_j^0 = \bigcup_{t\in{\cal T}_j } \left\{(i,j): \ \  i\in\Vc_{t}^{\emptyset}\cup\Vc_{t}^{-, j} \right\}\
 \end{equation}
where ${\cal T}_j$ is the set of pilot slots on which the $j$-th (detected) device has been detected by at least one sector in the network.

Secondly, the knowledge of the BS geographical locations can be exploited to determine some additional weak links  across the network.  For  two BSs which are {\em sufficiently} far apart, no device can have strong links {\em simultaneously} to {\em both} BSs. 
Let $d_{\rm thr}$ denote the distance beyond which two BSs are sufficiently far from one another so that no device can simultaneously have strong links to both BSs.\footnote{The distance $d_{\rm thr}$ can be experimentally obtained. Note that with this approach 
a tiny fraction of the ``known'' entries of $\Cm$ are misclassified, that is, $\widehat\Cm_{i,j}=0$, while $\Cm_{i,j}=1$. Hence, although, in principle, we should have $P_{\rm  D, max}=1$, in our experiments $P_{\rm  D, max}$ is slightly less than 1.}
Using knowledge of the distances between any pair of BSs in the network,  
the following set of entries (all along the $j$-th column) of $\Cm$ are also known to be zero:
\begin{equation}
\widecheck \Omega_j^0 = \bigcup_{i \in \Vc^j } \left\{(i',j): \ \  i'\in\Uc^{i}\right\},\
 \end{equation}
where $\Vc^{j}$ denotes the BS sectors that  have detected the $j$-th device, 
and where $\Uc^i$ denotes the set of all BS sectors whose distance from BS sector $i$ exceeds $d_{\rm thr}$. 

In summary, the set of all  known zero entries of $\Cm$ is given by $\Omega^0= \bigcup_{j=1}^{|\Dc|}\Omega^0_j$ , where  $\Omega^0_j = \widetilde \Omega^0_j \cup \widecheck \Omega^0_j$. 

\subsection{Matrix Completion Method}
\label{matrix-completion}

In this section we consider an online method for improving upon  the strong-link prediction performance provided by the baseline scheme.  The online method makes use of the intuitive spatial-consistency assumption that there is an underlying low-dimensional subspace describing the links between BS sectors and devices in the network.  Consequently, it uses the set of strong and weak links over the set $\Omega$ of (BS-sector, device) pairs (that are available to the central controller via the RA reports) as data to determine the ``best'' low-dimensional model and uses this model to  predict and classify links  in the complement set $\widebar\Omega$ of (BS-sector, device) pairs.

The online methods we consider fall within the class of matrix completion problems, which have recently gained attention in a broad range of applications. One important application is found in recommendation systems, and includes the ``Netflix Challenge'', according to which, movie recommendations are made to customers, based on their previous ratings and other-user ratings \cite{netflix-winner-2009}. At its core, this problem amounts to estimating the missing entries of a rating matrix where the matrix dimensions are the number of movies and the number of customers. While some entries of the matrix are already known, as some customers have rated some of the movies, rating estimates are required for  some movie-customer pairs. Since the  Netflix Challenge, various solutions have been devised, which take into account various practical aspects of the problem \cite{amazon-ieee2017}.  Recommendation systems research is very active, with many state-of-the-art works exploiting deep learning algorithms. 
As a proof of concept, in this paper we exploit a vanilla  solution based on matrix factorization (collaborative filtering with latent factor models) \cite{netflix-winner-2009}.
However, more advanced techniques can be also used and can, in principle, yield additional performance~benefits. 

In adapting the matrix completion problem to our setting,  we  rely on the following analogy: BS sectors are analogous to movies, devices are analogous to customers, and link gains between devices and sectors are analogous to movie ratings.

The problem of interest can be readily mapped into an intermediate matrix completion problem by noting that: (i) the central unit has available $\Gm_{i,j} $ for   $(i,j)\in\Omega^1$; (ii) the fact that $\Gm_{i,j}< \Gamma$  for $(i,j)\in\Omega^0$. As a result, we can formulate the following matrix completion problem:
\begin{eqnarray}
\label{matrix-completion-problem}
\min_{\displaystyle \Xm, \Thetam}&\!\!&
\sum_{(i,j)\in\Omega^1}\!\!\!|\Gm_{i,j}-\widecheck\Gm_{i,j}|^2
 \,+\!\! \!\! \sum_{(i,j)\in\Omega^0}\!\!\!|\Gamma^- -\widecheck\Gm_{i,j}|^2
\\
{\rm s.t.}&\!\!& \widecheck\Gm = \Thetam\Xm^T,\nonumber
\end{eqnarray}
for some $\Gamma^-$ appropriately chosen to optimize performance.\footnote{Practical aspects of the optimization problem, regarding rank selection, regularization etc. are discussed in detail in Appendix \ref{app-opt}.}

The solution $(\Xm^*, \Thetam^*)$ to  (\ref{matrix-completion-problem})  can be used to classify links {\em not} in $\Omega$. Indeed, by letting   $\widehat\Gm^* = \Thetam^*\Xm^{*T}$,  a hypothesis test can be employed for all $(i,j) \notin\Omega$, of the form 
\begin{equation}
\widehat\Cm_{i,j}=\begin{cases}
1& \text{if}\; \widehat\Gm^*_{i,j}\geq\beta\\\
0& \text{otherwise}
\end{cases},
\end{equation}
whereby $\beta$ is chosen so that the false-alarm probability does  not exceed a pre-assigned value $P_F$. 
\begin{figure}
\centering
\includegraphics[width=6.5cm, height = 2.5cm]{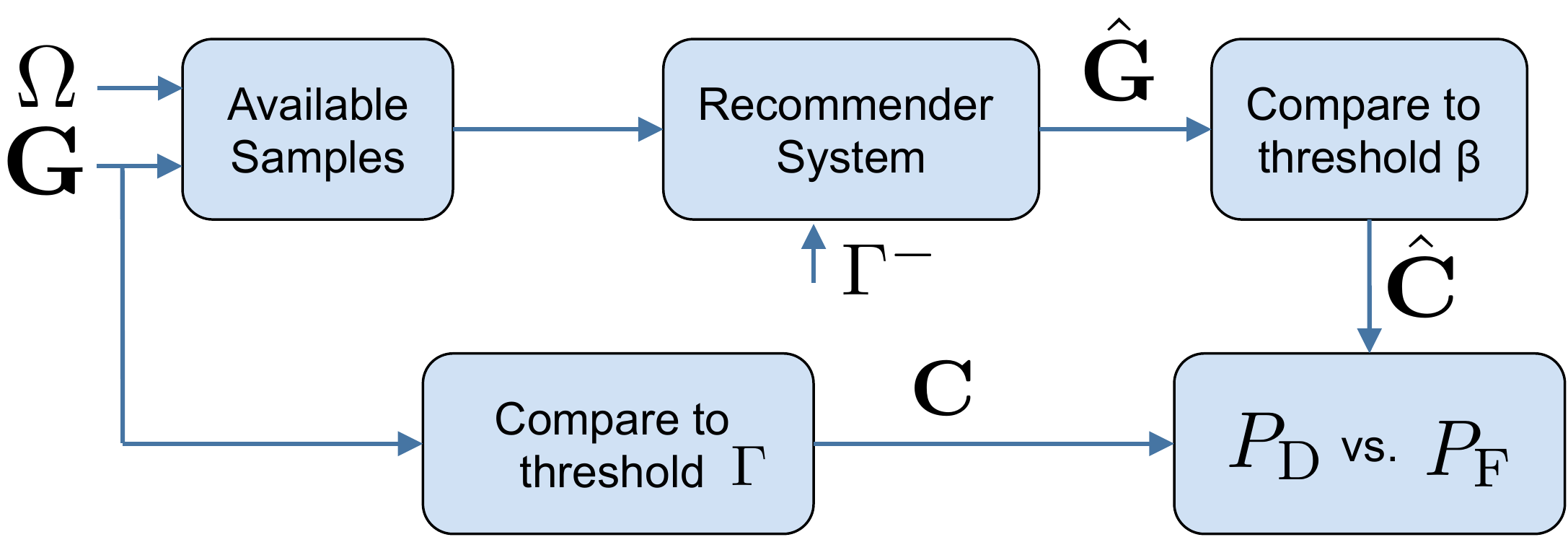}
\caption{Block diagram for matrix completion approach for online method}
\label{lambda1}
\vspace{-0.4cm}
\end{figure}
%
%
%
%
Varying $\beta$ between $\min(\widehat\Gm)$ and $\max(\widehat\Gm)$ can control the detection and false alarm rates, similar to the way the value of $\alpha$ controls these rates in the randomized baseline decision scheme.  A  block diagram for the proposed online-method  is shown  in Fig. \ref{lambda1}.

%

\section{Numerical Results}
\label{Numerical-Results}
In this section, we compare the performance of the matrix completion based sector-device link classification  against that of the baseline scheme in terms of  detection and false alarm rates as defined in Sec. \ref{link-detection}.

\begin{figure}
\centering
\includegraphics[width=7cm, height = 5cm]{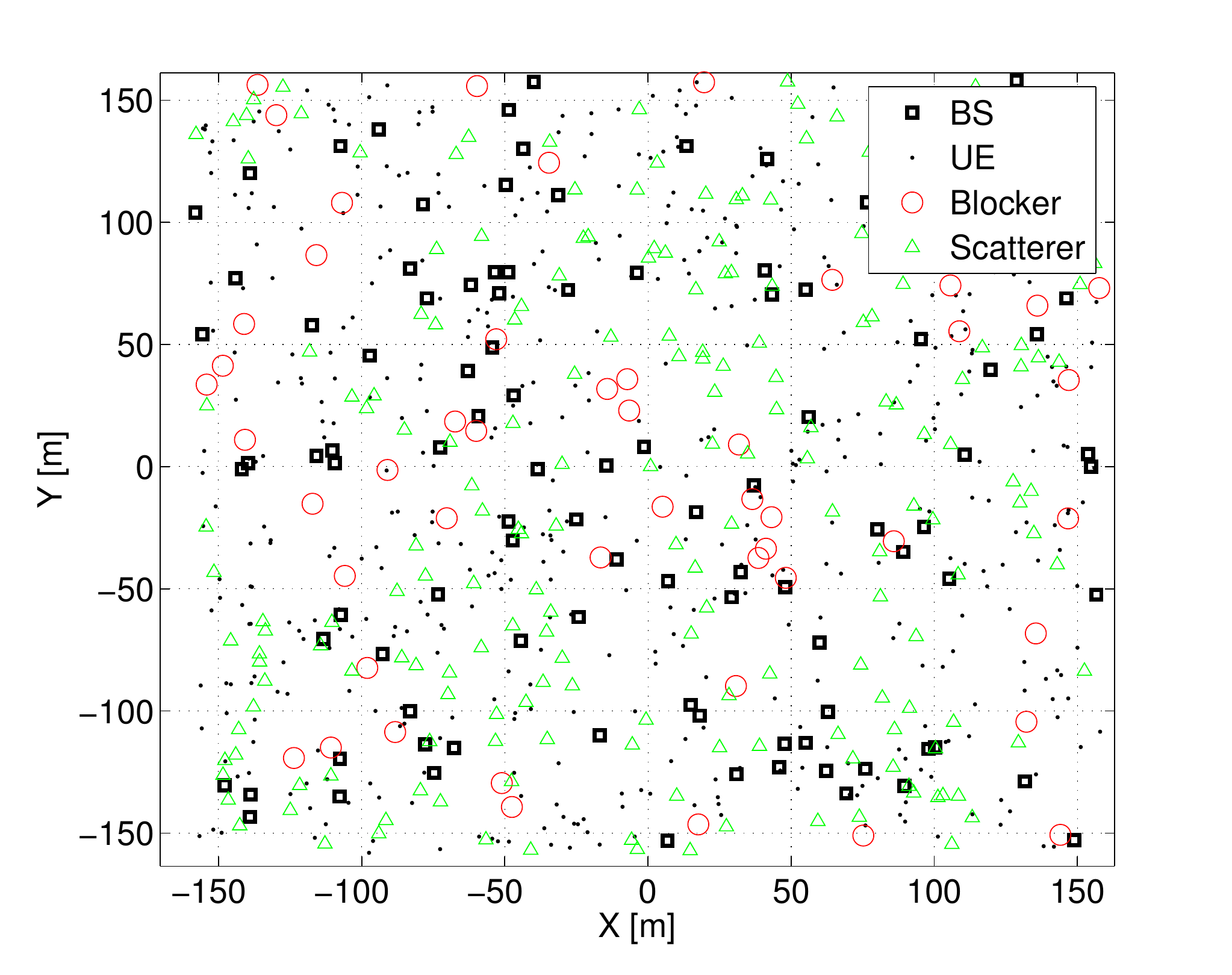}
\vspace{-0.2cm}
\caption{Sample layout including BSs, scatterers, blockers and users.}
\label{fig:BS-connectivity}

\end{figure}

We consider a network layout involving  $B = 100$ BSs  and large numbers of devices, both uniformly distributed over 
  a square geographical area of size $316 \,{\rm m} \times 316\, {\rm m}$.  Each BS has $S = 4$ sectors. The four sector orientation is chosen randomly and independently per BS.  Fig. \ref{fig:BS-connectivity} shows such a sample layout: BSs are shown with squares and devices are shown with dots (only $500$ devices are shown for representational clarity). 

\begin{figure}
\centering
\includegraphics[width=6.2cm, height = 4.5cm]{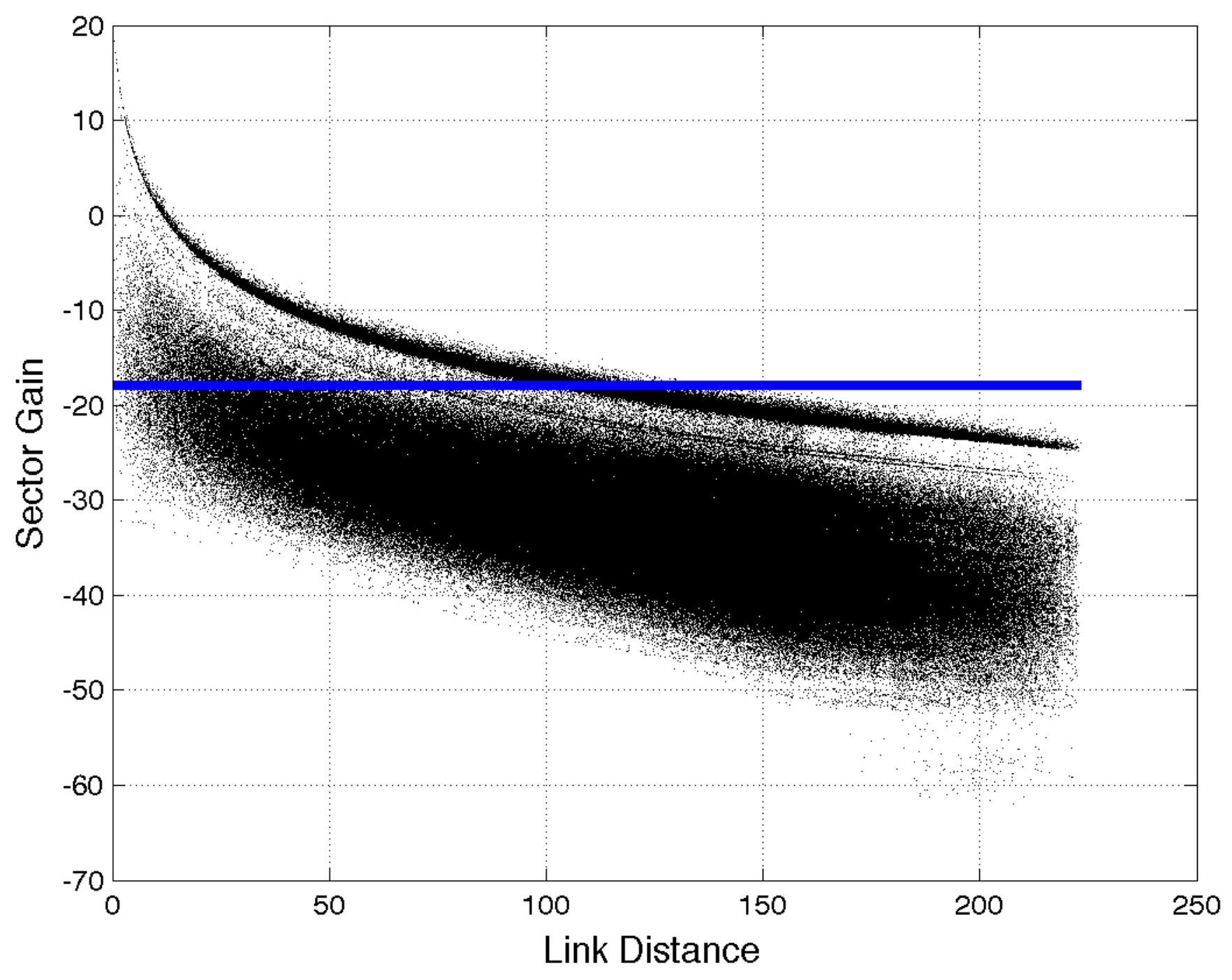}
\caption{Pathloss between RRH sectors and devices as a function of distance.}
\label{fig:Gain-distance profile}
\vspace{-0.4cm}
\end{figure}

The link strengths between BS-sectors and devices are determined via a generalization of the one-reflection pathloss model in \cite{bursalioglu2016novel} (described in detail Appendix \ref{app-cha}.
Fig. \ref{fig:Gain-distance profile} provides a plot of the strengths of all the links between the $400$ BS-sectors and the $500$  devices in Fig. \ref{fig:BS-connectivity} as a function of link distance. The horizontal line in the figure, set at $\Gamma = -18$ dB, reflects the a priori network-defined threshold, separating strong links from weak ones. As the figure reveals, links at similar distances can yield a broad range of link strengths. At the same time, the probability that a given link is strong decreases as a function of the distance. 

Our comparisons  are based on the use of the RA protocol in Sec. \ref{specific-RA}, and
are tuned such that an arrival rate of $\lambda_{\rm in} = 500$ devices can be accommodated with an expected delay requirement of $1/\rho$, where $\rho = 0.99$. 
The RA-block within each frame is assumed to have an overhead of $T = \theta \lambda_{\rm in}$  pilot dimensions, where $\theta=T/\lambda_{\rm in}$ denotes the {\em relative} overheads of the RA scheme.  Given $\lambda_{\rm in} = 500$ and $\rho = 0.99$, the target operating point is $\lambda_{\rm out}^{\rm tar.} = 500$ and $\lambda_{\rm ra}^{\rm tar.} = 506$. With this configuration 99\% of the active devices are detected on average during the RA block of a single frame.  Implicit in this operating-point specification, however, is the choice of the $p$ parameter, which, given a particular $T$ overhead, guarantees  $\lambda_{\rm out} = \lambda_{\rm out}^{\rm tar.} = \lambda_{\rm in}$, with delay requirements $1/\rho$.  
  
In principle the value of the $p$ parameter can be determined and tuned via simulations. In particular, as discussed in Sec. \ref{sec:RA}, in order to stabilize the queues and, in addition, meet the delay requirements, the scheme should guarantee the following:
\begin{equation}
\label{lambda_constraints}
\lambda_{\rm out} = \lambda_{\rm in}\;\; \text {and }\;\; \lambda_{\rm out}\geq \rho\lambda_{\rm ra}.
\end{equation}
\vspace{-0.05cm}
Hence, given $\lambda_{\rm in}$ and $T$, we can estimate $\lambda_{\rm out}$ and $\lambda_{\rm ra}$ as a function of $p$ via simulations,  and subsequently use the $p$ value that satisfies the constraints in (\ref{lambda_constraints}).  


As an alternative to extensive simulations, we  develop a simple method for predicting the value of $p$ yielding (\ref{lambda_constraints}) and investigate its match in the parameter range of interest based on simulations. The method is based on modeling the device detection process at each sector. This model assumes that, at any given slot $t$ out of the $T$ slots in the RA training block, each  active user not yet detected has a probability equal to some value $q$ to be detected in that slot (independent of all other users). Based on this rudimentary model, we can derive the following rule-of-thumb formula between $\lambda_{\rm out}^{\rm model}$ and $ \lambda_{\rm ra}$:
\begin{equation}
\label{model}
\lambda_{\rm out}^{\rm model}(q,T) = \lambda_{\rm ra}\left[1-(1-q)^{T}\right].
\end{equation}

\begin{figure}
\centering
\includegraphics[width=6cm, height = 4.8cm]{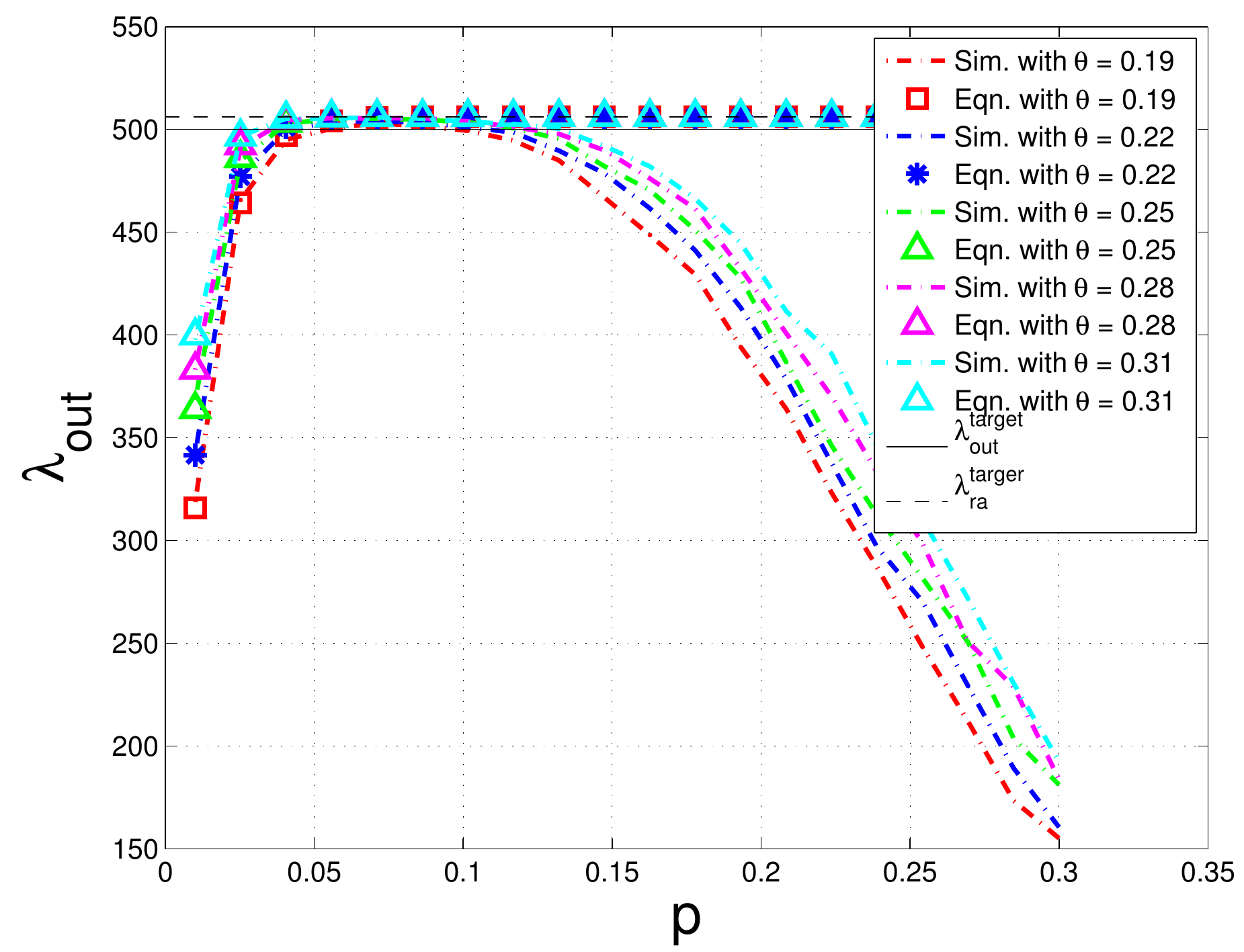}
\caption{$\lambda_{\rm out}$ vs. $p$ curves for various  RA overheads}
\label{lambda2}
\vspace{-0.4cm}
\end{figure}

For the network in Fig.~\ref{fig:BS-connectivity}, Fig. \ref{lambda2} shows a simulation-based  evaluation of the efficacy of  the formula in (\ref{model})  using $q=p$ and $\lambda_{\rm ra} = \lambda_{\rm ra}^{\rm tar.} = 506$, as a function of $\theta=T/\lambda_{\rm in}$ and $p$. As the figure reveals, the $\lambda_{\rm out}$ value predicted by the model matches simulations especially well for the values of $p$ that maximize $\lambda_{\rm out}$ and the values of $\theta$ that can reach $\lambda_{\rm out}$.  As a result,
for any given operating point of interest given  by a triplet of ($\rho$, $\lambda_{\rm in}$,  $\theta$) values, we set 
$p = p^*(\theta)=1-(1-\rho)^{\frac{1}{\theta\lambda_{\rm in}}}$.


%

\begin{figure}
\centering
\includegraphics[width=6.5cm, height = 5.5cm]{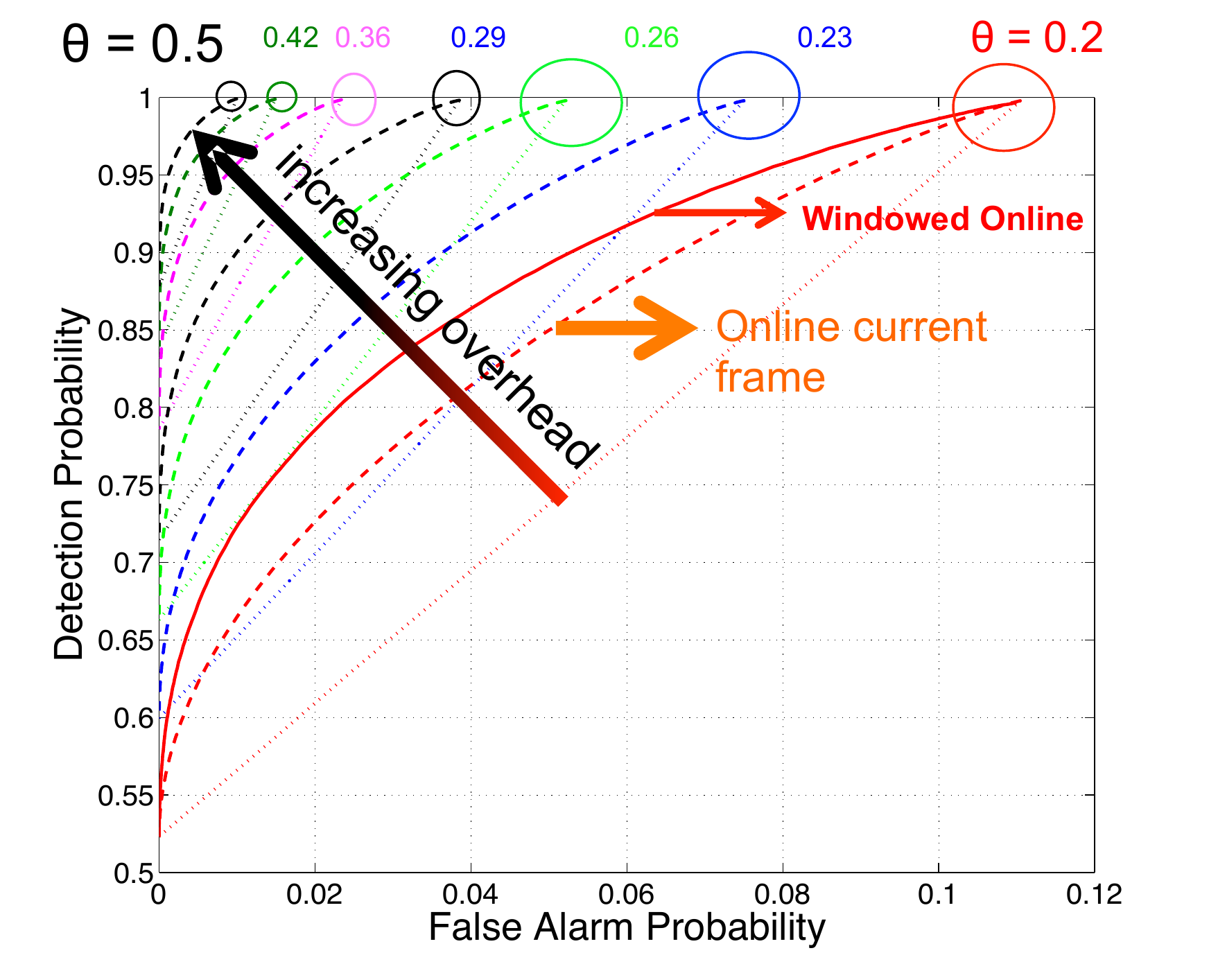}
\vspace*{-0.4cm}
\caption{Detection probability vs. False Alarm for various RA overheads}
\label{FA_PD}
\end{figure}

Fig. \ref{FA_PD} shows the probability of detection vs. false alarm of strong links for various $\theta$ values. For each $\theta$ value, the dotted  line  depicts the performance of the baseline scheme, while the associated dashed curve corresponds to the matrix-completion online method. The red solid curve in the figure shows the performance of the matrix-completion method at $\theta=0.2$, operating over a window of the two last frames (current and most recent past frame) in predicting the links of current-frame active devices. As the figure reveals, the online methods yield sizable advantages in predicting strong links. Furthermore, the performance improves when the online method is applied over a window of the two most recent frames. 
%
%
%
%

%
%

\vspace{-0.2cm}

\section {Conclusion}
\label{sec:conclusion}

We have studied the efficacy of matrix-completion algorithms in classifying the strengths of links between detected devices and RRH sectors, based on C-RAN RA observations collected across a wide-area network.  Our analysis and simulations reveal that such purely data-driven schemes can significantly  improve link classification. More important, link classification becomes more accurate when the proposed methods are applied over windows of consecutive RA frames, allowing the algorithm to work on  larger numbers of devices.  

Our work indicates the existence of an efficient representation for the collection of link-strength device vectors  in terms of a low-dimensional ``feature'' subspace. One can foresee that, with lots of IoT devices, these feature vectors  can be accurately learned and subsequently used to make link predictions per device. Such techniques are left for future work.


\vspace{-0.1cm}


%

\appendix
{
\subsection{Notes on the practical issues of matrix completion problem}
\label{app-opt}

For simplicity, we let $\Gamma^-=\Gamma$ in (\ref{matrix-completion-problem}) and add regularization terms:

\begin{eqnarray}
\min_{{\Xm},{\Thetam}}&\sum_{(i,j)\in\Omega^1}|\Gm_{i,j}-\widecheck\Gm_{i,j}|^2+\;\sum_{(i,j)\in\Omega^0}|\Gamma -\widecheck\Gm_{i,j}|^2\nonumber\\
 &+\;\;\lambda(\|\mathbf{\Theta}\|_F^2+\|\mathbf{X}\|_F^2) \label{Eq.prob}\\
{\rm s.t.}& \check\Gm = \Thetam\Xm^T,\nonumber
\end{eqnarray}
{where $\|\cdot\|_F$ denotes the Frobenius norm. In (\ref{Eq.prob}), the first two terms indicate the sum square of estimation error, while the third term represents the regularization factor to reduce the overfitting effect \cite{Ng-Coursera-ML}. Although (\ref{Eq.prob}) is not  jointly convex with $\mathbf{\Theta}$ and $\mathbf{X}$, it is a convex quadratic function of one variable if we fix others. Therefore, a simple block descent algorithm can be considered to achieve a local optimal solution:
\begin{itemize}
\item
Following the first order optimality condition, find the update equations of $\mathbf{\Theta}$ and $\mathbf{X}$, respectively.
\item
With a random initialization, iteratively evaluate variables by update equations until the estimation error converges or the maximum number of iterations is reached. 
\end{itemize}}

{To achieve a sub-optimal solution of the matrix factorization problem and estimate unknown links, we fix the parameter settings in Tab. \ref{Tab.paraSet} empirically,
\begin{table}
\caption{Parameter settings of collaborative filtering}
\begin{center}
\begin{tabular}{c|c}
\hline
$\lambda$ & $20$\\ 
 \hline
$r$ & $200$\\  
 \hline
Step size of gradient descent & $5\times 10^{-5}$\\
 \hline
Maximum no. of iterations & $1000$\\
\hline
$\epsilon$ & $10^{-2}$\\
\hline
\end{tabular}
 \label{Tab.paraSet}
\end{center}
\end{table}
where $\epsilon$ is the parameter to lower bound the normalized sum square of estimation error over the training data (known link entries):

\begin{align}
\frac{\sum_{(i,j)\in\Omega^1}|\Gm_{i,j}-\check\Gm_{i,j}|^2+\sum_{(i,j)\in\Omega^0}|\Gamma-\check\Gm_{i,j}|^2}{\sum_{(i,j)\in\Omega^1}|\Gm_{i,j}|^2+\sum_{(i,j)\in\Omega^0}|\Gamma|^2}\geq \epsilon.
\label{Eq.errorBound}
\end{align}

Therefore, either (\ref{Eq.errorBound}) is satisfied or the number of iterations reaches its maximum, the algorithm stops and outputs the result of matrix factorization. }
}
\subsection{ Channel model}
\label{app-cha}
We consider a radio access network (RAN) with $B$ RRHs and $K$ UEs. Each RRH is equipped with an M-element uniform linear array (ULA), while each UE is equipped with a single antenna. With the assumption of the block fading channel \cite{wc_book_tse},  the channel response vector between the $j$-th RRH and the $k$-th UE, i.e., $\mathbf{h}_{jk}\in\mathbb{C}^{M\times1}$, can be represented as
\begin{align}
\mathbf{h}_{jk}=\sum_{n=1}^{N_{jk}}\sqrt{L_{jk,n}}\beta_{jk,n}\mathbf{a}(\theta^{\rm D}_{jk,n})e^{-j2\pi\tau_{jk,n}f},
\label{Eq.channel}
\end{align} 
which is a superposition of MPCs. In (\ref{Eq.channel}), $f$ is the carrier frequency, $N_{jk}$ indicates the number of MPCs between $\text{RRH}_j$ and $\text{UE}_k$, and $L_{jk,n}$ is the large scale loss for the $n$-th MPC of link $\text{RRH}_j$-to-$\text{UE}_k$. $\beta\sim\mathcal{CN}(0,1)$ reflects the small scale fading following complex Gaussian distribution, and $\tau$ is the propagation time of MPC. $\theta^{\rm D}$ represents the direction of departure (DOD) from the perspective of RRH, and $\mathbf{a}(\theta^{\rm D})$ is the steering vector with respect to $\theta^{\rm D}$:
\begin{align}
\mathbf{a}(\theta^{\rm D})\triangleq[1,e^{-j2\pi\frac{d^{\rm ant.}\cos{\theta^{\rm D}}}{\lambda}},...,e^{-j2\pi(M-1)\frac{d^{\rm ant.}\cos{\theta^{\rm D}}}{\lambda}}]^T,
\end{align} 
where $d^{\rm ant.}$ and $\lambda$ indicate the antenna spacing and wavelength, respectively. Note that we only consider the two-dimensional channel model in the azimuth domain in (\ref{Eq.channel}), but it can be easily generalized to incorporate the elevation domain.  

Within the stationarity region of the second order channel statistics, long-term channel parameters, including DODs of MPCs $[\theta^{\rm D}_{jk,n}]$ and large scale loss $[L_{jk,n}]$ remain approximately the same for a relatively long period of time, which can be equivalent to hundreds of coherence blocks \cite{adhikary2013joint}. On the other hand, small scale variables, e.g., $[\beta_{jk,n}]$, remain the same within the coherence time and coherence bandwidth, but vary independently across different blocks. Therefore, based on the assumption of uncorrelated scattering, similar to \cite{adhikary-mmWave-jsac}, we can achieve the close-form expression of channel covariance as a function of long-term large scale parameters:
\begin{align}
\mathbf{K}_{jk}\triangleq \mathbb{E}[\mathbf{h}_{jk}\mathbf{h}_{jk}^\dag]=\sum_{n=1}^{N_{jk}}L_{jk,n}\mathbf{a}(\theta^{\rm D}_{jk,n})\mathbf{a}^\dag(\theta^D_{jk,n}).
\label{Eq.Cov}
\end{align}     

We consider the DFT based virtual sectorization within each RRH, where each sector is formed by spatially filtering the ULA signal through a contiguous set of sector-specific DFT beams. Define $\mathbf{\Omega}_M$ as the $M\times M$ normalized DFT matrix (each column has unit norm), we acquire the average sector gain between the $i$-th sector of $\text{RRH}_j$ and $\text{UE}_k$ as 
\begin{align}
\mathbf{G}(i+(j-1)S,k)=\text{tr}(\mathbf{\Omega}_M^\dag(\mathcal{S}_{ji},:)\mathbf{K}_{jk}\mathbf{\Omega}_M(\mathcal{S}_{ji},:)), \forall i, j, k,
\label{Eq.sectorGainMat}
\end{align} 
where $\mathbf{G}\in\mathbb{R}_+^{BS\times K}$ indicates the sector gain matrix, $S$ is the number of sectors per RRH, and $\mathcal{S}_{ji}$ denotes set of DFT beam indices for the $i$-th sector of $\text{RRH}_j$. Splitting all DFT beam tones equally among sectors of an RRH, we have $\mathcal{S}_{ji}\triangleq\{x|1+(i-1)\frac{M}{S}\leq x\leq i\frac{M}{S}, x\in\mathbb{Z}_+\}$. 

To generate synthetic channel profiles, we implement a geometric stochastic channel model (GSCM) \cite{bursalioglu2016novel}, which captures the propagation characteristics of direct LOS path and non-direct reflected path. Since MPCs interacting with multiple objects will usually suffer severe propagation loss, especially at high frequencies \cite{molischbook}, we only consider the non-direct path with single-bounce scattering. Blockage effects \cite{bai2014analysis} are also incorporated by adding circle-shaped blockers so that the existence of direct path is determined geometrically. 

For the attenuation of the direct path, we utilize a smooth transition function \cite{constantine2005antenna}:
\begin{align}
g(d)=(1+\frac{d}{\epsilon})^{-\alpha},
\label{Eq.LOS}
\end{align}  
where $d\geq 0$ denotes the separation distance between two link ends, and $\epsilon$ and $\alpha$ indicate the breakpoint distance and pathloss exponent, respectively. Regarding to the non-direct paths, we combine the separate loss from RRH-to-scatterer and scatterer-to-UE:
\begin{align}
f(d_{uz},d_{zr})=ag(d_{uz})g(d_{zr}),
\label{Eq.NLOS}
\end{align}
which is commonly used for modeling reflected path attenuation \cite{wc_book_tse}. In (\ref{Eq.NLOS}), $a\leq 1$ is the attenuation coefficient of NLOS path, and $d_{uz}$ and $d_{zr}$ denote the distance of UE -to-scatterer and scattere-to-RRH, respectively. More discussions on the validity of (\ref{Eq.NLOS}) can be found in \cite{bursalioglu2016novel}. 

In conclusion, we briefly summarize the procedures of channel modeling below:
\begin{itemize}
\item
In the operating area, randomly drop RRHs, UEs, scatterers, and blockers satisfying separate Poisson point process (PPP).
\item
For the direct path of a particular link, if it is blocked by any blocker, there is no LOS path for that link.
\item
Calculate the attenuation $[L_{jk,n}]$ of LOS path and NLOS path according to (\ref{Eq.LOS}) and (\ref{Eq.NLOS}), respectively.  
\item
Remove weak NLOS paths, whose attenuation is below than a threshold $\gamma_\text{path}$. 
\end{itemize}   
Since scatterers of a link are usually not too far away from either the RRH or the UE, the last step will eliminate non-realistic paths, and maintain the set of effective scatterer for every link, respectively. Note that this GSCM model provides an implementation of the “common scatterer” concept used, e.g., in COST 2100 \cite{liu2012cost}.

With the synthetic channel profiles, we can extract the angle information $[\theta^D_{jk,n}]$ following the geometric relations of RRH, scatterer, and UE. Combining with the knowledge of large scale loss $[L_{jk,n}]$, we acquire the sector gain matrix $\mathbf{G}$ according to (\ref{Eq.Cov}) and (\ref{Eq.sectorGainMat}).

 \bibliographystyle{IEEEtran}
\bibliography{IEEEabrv,refs}
\end{document}